\documentclass[11pt,a4paper]{article}

\usepackage{mathpple}
\usepackage{graphicx}
\usepackage{url}
\usepackage{paralist}

\usepackage{geometry}
\usepackage{fancyhdr}
\begin{document}

\thispagestyle{plain}

\noindent \textbf{Preprint of:}\\
Dmitri K. Gramotnev and Timo A. Nieminen\\
``Non-steady-state double-resonant extremely asymmetrical
        scattering of waves in periodic gratings''\\
\textit{Physics Letters A} \textbf{310}, 214--222 (2003)

\hrulefill

\begin{center}

\textbf{\LARGE
Non-steady-state double-resonant extremely asymmetrical
        scattering of waves in periodic gratings
}

{\Large
Dmitri K. Gramotnev$^1$ and Timo A. Nieminen$^2$}

{}$^1$Applied Optics Program, School of Physical and Chemical Sciences,
Queensland University of Technology, GPO Box 2434, Brisbane,
QLD 4001, Australia; e-mail: d.gramotnev@qut.edu.au

{}$^2$Centre for Biophotonics and Laser Science, Department of Physics,
The University of Queensland, Brisbane, Qld 4072, Australia

\vspace{1cm}

\begin{minipage}{0.8\columnwidth}
\section*{Abstract}
Double-resonant extremely asymmetrical scattering
(DEAS) is a strongly resonant type of Bragg scattering in two joint or
separated uniform gratings with different phases. It
is characterised by a very strong increase of the
scattered and incident wave
amplitudes inside and between the gratings at a
resonant phase shift between the gratings. DEAS
is realised when the first
diffracted order satisfying the Bragg condition
propagates parallel to the grating boundaries,
and the joint or separated gratings
interact by means of the diffractional divergence
of the scattered waves from one grating into another.
This Letter develops a
theory of non-steady-state DEAS of bulk TE electromagnetic
waves in holographic gratings, and investigates the process of
relaxation of the incident and scattered wave amplitudes to
their steady-state values inside and outside the gratings. Typical
relaxation times are determined. Physical explanation of the
predicted effects is presented.
\end{minipage}

\end{center}

\section{Introduction}

Double-resonant extremely asymmetrical scattering
(DEAS) [1--4] is a strongly resonant wave effect
in slanted, strip-like, periodic gratings with the
scattered wave propagating parallel to the grating
boundaries. DEAS occurs in a non-uniform grating that
consist of two strip-like joint [1,2,4] or separated [3]
uniform gratings with different phases. It is characterised
by a unique combination of two simultaneous
resonances---one with respect to frequency, and the
other with respect to phase shift between the gratings
[1--4]. The resonance with respect to frequency is
typical for all types of scattering in the geometry of
extremely asymmetrical scattering (EAS), and results in
a strong increase of the scattered wave amplitude
inside and outside the grating [5--8]. The second
resonance with respect to phase shift between the joint or
separated gratings is typical only for DEAS and
results in a strong resonant maximum that occurs on the
background of already resonantly large scattered wave
amplitude (typical for EAS). The resonant phase shift
is usually close to $\pi$, and the resultant scattered wave
amplitudes may well exceed tens or hundreds of the
amplitude of the incident wave at the front boundary
[1--4].

   It is important that not only the scattered wave
amplitude, but also the amplitude of the incident wave
(the 0th diffracted order) near the interface between
the joint gratings, or in the gap between the two
gratings experiences a strong resonant increase at the
resonant phase shift [1--4].

   It has also been demonstrated that strong DEAS
occurs only in a grating of width that is smaller than
a determined critical width [1,2]. Physically, half of
the critical width was shown to equal the distance
within which the scattered wave can be spread across
the grating by means of diffractional divergence,
before being rescattered by the grating [1,2]. Thus,
diffractional divergence of the scattered waves from
one of the uniform gratings into another was shown
to be one of the main physical reasons for DEAS
[1--3]. Two simple methods for the determination of
the critical width were suggested and justified [1,2].

    On the basis of understanding the role of diffractional
divergence of the scattered wave for DEAS, an
efficient approximate method of analysis of this type
of scattering was developed [1--3]. The main advantage
of this method is that it is immediately applicable
for the analysis of DEAS of all types of waves
in various kinds of periodic gratings, including guided
and surface electromagnetic and acoustic waves. This
method also provides an excellent insight into the
physical processes in uniform and non-uniform gratings
in the geometry of EAS. Comparison of the
approximate and rigorous theories of DEAS [4]
demonstrates their good agreement in gratings with small
amplitude, if the DEAS resonance is not too strong.

    As with any other strongly resonant effect, DEAS
must be characterised by a significant time of
relaxation to the analysed steady-state regime of
scattering. The higher the resonance, the larger the
relaxation time for the particular grating. On the other
hand, the larger the relaxation time, the larger the
distance that the scattered wave must travel along the
grating boundaries before its amplitude reaches the
steady-state values. Thus the aperture of the incident
beam must also increase proportionally. Therefore,
each time of relaxation is associated with a particular
critical aperture of the incident beam that is required
for the steady-state regime of DEAS to be achieved
(see also [6--9]). As a result, not all DEAS resonances
can be readily achieved in practice. In some cases,
the critical aperture of the incident beam (and thus
the length of the grating) may be unreasonably large.
Therefore, the accurate knowledge of relaxation times
and critical apertures of the incident beam for different
gratings with DEAS is essential for the experimental
observation and practical application of this strongly
resonant wave effect.

   Consistent approximate and rigorous theories of
non-steady-state EAS, based on the Fourier analysis
of the incident pulse that is `switched on' at some
moment of time, have recently been developed in
uniform gratings [9]. This approach has resulted not
only in accurate determination of the relaxation times
for any point inside and outside the grating, but also
in the detailed analysis of non-steady-state EAS in
uniform gratings. Though the same approach must be
applicable for the analysis of non-steady-state DEAS,
this has not been done so far.

   Therefore, the aim of this Letter is to present
the detailed analysis of non-steady-state DEAS in
two joint or separated uniform gratings with a phase
shift. Temporal evolution of the scattered and incident
wave amplitudes inside and outside the grating will
be investigated numerically. Accurate relaxation times
and critical apertures of the incident beam will be
determined. Approximate and rigorous approaches
will be discussed and compared.

\section{Structure and methods of analysis}

\begin{figure}[!htb]
\centerline{\includegraphics[width=0.5\columnwidth]{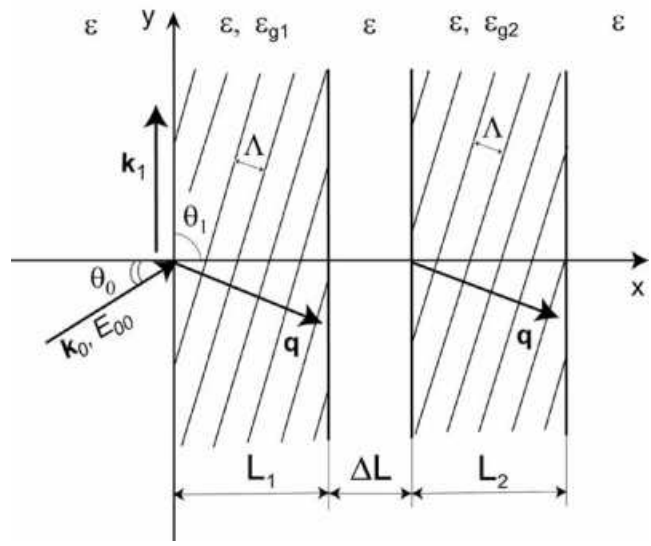}}
\caption{The structure with DEAS in two uniform slanted gratings
of widths $L_1$ and $L_2$ , separated by a gap of width $\Delta L$. The
incident rectangular pulse: central frequency $\omega_0$, the corresponding
wavevector $\mathbf{k}_0$, amplitude of the electric field in the pulse
$E_{00}$.}
\end{figure}

   Consider an isotropic medium with two slab-like,
uniform, holographic grating with sinusoidal modulation
of the dielectric permittivity (Fig. 1):
\begin{eqnarray}
\epsilon_s & = & \epsilon + \epsilon_{g1}
\exp(\mathrm{i}q_xx + \mathrm{i}q_yy)  + \epsilon_{g1}^\ast
\exp(-\mathrm{i}q_xx - \mathrm{i}q_yy) \nonumber \\
& & \textrm{if} \,\,\,\, 0<x<L_1, \nonumber\\
\epsilon_s & = & \epsilon + \epsilon_{g2}
\exp(\mathrm{i}q_xx + \mathrm{i}q_yy)  + \epsilon_{g2}^\ast
\exp(-\mathrm{i}q_xx - \mathrm{i}q_yy) \nonumber \\
& & \textrm{if} \,\,\,\, L_1 + \Delta L<x<L_1+\Delta L + L_2, \nonumber\\
\epsilon_s = \epsilon \,\,\,\, \textrm{elsewhere},
\end{eqnarray}
where $L_{1,2}$ are the widths of the two gratings,
$\epsilon_{g1,2}$
are the grating amplitudes, $\Delta L$ is the width of the gap
between the uniform gratings---see Fig. 1 ($\Delta L$ may be
zero, and then we have two joint gratings), the mean
dielectric permittivity, $\epsilon$, is the same everywhere in
the structure, $q_x$ and $q_y$ are the $x$- and $y$-components
of the reciprocal lattice vector $\mathbf{q}$ that is the same for
both the gratings, $q = 2\pi/\Lambda$, $\Lambda$ is the period of the
gratings, the coordinate system is shown in Fig. 1.
The two gratings have different phases:
$\phi = \mathrm{arg}(\epsilon_{g2}) - \mathrm{arg}(\epsilon_{g1})$
is the phase shift between the gratings. There
is no dissipation in the medium ($\epsilon$ is real and positive),
and the structure is infinite along the $y$- and $z$-axes.

    Non-steady-state DEAS in the this structure occurs
when the incident wave is switched on at some
moment of time (e.g., at $t = 0$). Then, both the incident
and scattered wave amplitudes inside and outside the
gratings evolve in time and gradually relax to their
steady-state values at $t = +\infty$. This occurs when an
infinitely long, sinusoidal, step-function incident pulse
is switched on at t = 0 at the front grating boundary.
The pulse has the amplitude of the electric field $E_{00}$,
infinite aperture along the $y$- and $z$-axes, and the angle
of incidence $\theta_0$ (non-conical geometry---Fig. 1).

   However, the numerical analysis of an infinitely
long pulse is inconvenient, since its Fourier transform
contains a  $\delta$-function. Therefore, the method of
analysis of non-steady-state EAS, developed in [9], has
been based on the consideration of a rectangular
sinusoidal incident pulse of finite (rather than infinite)
length. The same method is immediately applicable
for the approximate and rigorous analyses of DEAS
in the considered structures (Fig. 1). In this case, in
order to calculate non-steady-state amplitudes of the
incident and scattered waves in the structure at an
arbitrary moment of time $t = t_0$, we consider an
incident pulse (Fig. 1) of the time length $2t_0$ [9].
Paper [9] also assumed that at $t = 0$ the incident pulse
was switched on simultaneously (with the amplitude
$E_{00}$) everywhere inside the grating and in front of it
within the region $L - 2t_0c^{-1/2} \cos\theta_0 < x < L$, where
$L = L_1 + \Delta L+L_2$. That is, the process of propagation
of the pulse through the gratings was ignored (for a
detailed justification of this approximation see [9]). In
this Letter, we will compare this approximation with
the case when the incident wave is switched on at
$t = 0$ everywhere at the front grating boundary and
then propagates through the grating. It is important
that both these methods can be equally used for the
analysis of non-steady-state DEAS (see below).

    The incident rectangular pulse at $t = 0$ is expanded
into the Fourier integral with respect to time, and its
frequency spectrum is determined analytically. As a
result, the incident pulse is represented by a
superposition of an infinite number of plane waves having
different frequencies and amplitudes (determined by the
Fourier coefficient in the integral), and the same
angle of incidence $\theta_0$ (Fig. 1). Each of these plane waves
is regarded as an incident plane wave with the determined
amplitude at the front grating boundary. Steady-state
scattering of these waves is then analyzed by
means of the rigorous [4,10] or approximate (if
applicable) [1--3] theory of steady-state DEAS.

   The rigorous theory [4,10] is based on the enhanced
T-matrix algorithm [11,12]. As a result, we obtain
steady-state amplitudes of the 0th and $+1$ diffracted
orders inside and outside the gratings, corresponding
to each of the plane waves in the Fourier spectrum of
the incident pulse. Interference of the steady-state $+1$
diffracted orders at the moment of time $t = t_0$ (i.e.,
the inverse Fourier transform applied to these scattered
waves) gives the overall (non-steady-state) scattered
wave amplitude as a function of the $x$-coordinate [9].
Note that due to the geometry of the problem the non-
steady-state incident and scattered wave amplitudes
should not depend on the $y$-coordinate. Similarly,
applying the inverse Fourier transform to all the 0th
diffracted orders in the gratings, gives the overall (non-
steady-state) amplitude of the incident pulse in the
grating at $t = t_0$ as a function of the $x$-coordinate [9]
Note that in order to minimize numerical errors [9],
the inverse Fourier transform is taken at $t = t_0$, i.e., at
the middle of the incident pulse. Applying the same
procedure for different values of $t_0$ (i.e., for incident
pulses of different time length $2t_0$), we obtain the
time evolution of non-steady-state amplitudes of the
incident and scattered waves inside and outside the
gratings.

   Very similarly, for gratings with small amplitude,
the approximate theory of steady-state DEAS, based
on the allowance for the diffractional divergence
of the scattered wave [1--3], can be used instead
of the rigorous coupled wave analysis. However, in
the case of non-steady-state DEAS, the approximate
theory does not lead to simple analytical results (as
in papers [1--3]). Therefore, there is little reason to
use the approximate approach from the view-point
of simplification of the analysis. Nevertheless, the
approximate approach of non-steady-state DEAS is
important because it is immediately applicable for
the analysis of all types of waves in different kinds
of periodic gratings with small grating amplitude
(e.g., for guided modes in a slab with a corrugated
interface---see [2,3,7]).

   The described rigorous and approximate theories of
non-steady-state DEAS are applicable for all (not only
rectangular) shapes of the incident pulse, as well as
for an incident beam of finite aperture. However, for
beams of finite aperture, we should also use the spatial
Fourier analysis.

   Due to large time intervals required for the analysis
of non-steady-state DEAS, a logarithmic distribution
of time points has been used (thus, the fast Fourier
transform cannot be used for this analysis). The
calculations are carried out separately for each moment
of time $t_0$. Therefore, there is no accumulation of
numerical errors, and noticeable errors at small times
$\approx 10^{-13}$s do not affect the accuracy of the results at
larger time intervals (see above and [9]).

\section{Numerical results}

   Using the described numerical algorithm, non-
steady-state DEAS of bulk TE electromagnetic waves
in non-uniform holographic gratings with a phase shift
$\phi \ne 0$ has been analysed. The grating parameters are
as follows: $\epsilon = 5$,
$\epsilon_{g1} = |\epsilon_{g2}| = 5\times 10^{-3}$,
$\theta_0 = 45^\circ$,
and the wavelength in vacuum (corresponding to the
central frequency $\omega_0$ of the spectrum of the incident
rectangular pulse) $\lambda_0 = 1\mu$m. First, consider the case
with $\Delta L = 0$ (two joint uniform gratings). We also
assume that $L_1 = L_2$ (the effect of non-equal grating
widths of the joint gratings on EAS is considered
in [2]). The Bragg condition is assumed to be satisfied
precisely for the $+1$ diffracted order at $\omega = \omega_0$:
\begin{equation}
\mathbf{k}_1(\omega_0) = \mathbf{k}_0(\omega_0) - \mathbf{q}
\end{equation}
where $\mathbf{k}_0(\omega)$ are the frequency dependent wave
vectors of the plane waves in the Fourier integral of
the incident pulse, $\mathbf{k}_1(\omega)$ are the wave vectors of the
corresponding $+1$ diffracted orders (scattered waves),
$k_1(\omega_0) = k_0(\omega_0) = \omega_0\epsilon^{1/2}/c$
(i.e., the Bragg condition
is satisfied for the central frequency $\omega_0$),
$\mathbf{k}_1(\omega_0)$ is parallel to the grating
boundaries (Fig. 1), $c$ is the speed
of light. Note that if $\omega \ne \omega_0$, $\mathbf{k}_1(\omega)$
is not parallel to the grating boundaries [13].

\begin{figure}[!t]
\centerline{\includegraphics[width=0.6\columnwidth]{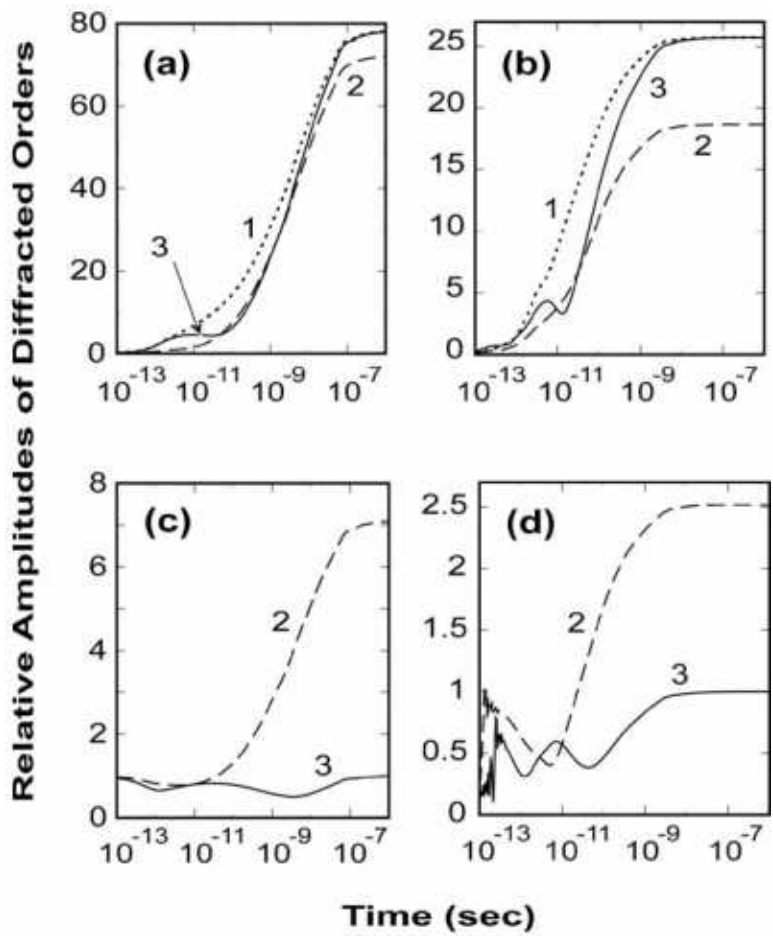}}
\caption{The typical time dependencies of the relative
non-steady-state amplitudes of (a), (b) the $+1$ diffracted order
(scattered wave) $|E_1/E_{00}|$, and (c), (d) the 0th diffracted order
(incident wave) $|E_0/E_{00}|$ in the holographic gratings with the
parameters: $\epsilon = 5$,
$\epsilon_{g1} = |\epsilon_{g2}| = 5\times 10^{-3}$,
$\theta_0 = 45^\circ$, the vacuum wavelength at $\lambda_0 = 1\mu$,
$\Delta L = 0$ (joint gratings), at different grating widths: (a),
(c) $L_1 = L_2 = 10\mu$m and
$\phi = \mathrm{arg}(\epsilon_{g2} = \phi_r \approx 191.2^\circ$,
(b), (d) $L_1 = L_2 = 15\mu$m and $\phi = \phi_r \approx 205.2^\circ$.
Dotted curves---front boundary ($x = 0$), dashed
curves---middle of the grating, i.e., the boundary between the
gratings ($x = L_1$), solid curves---rear boundary ($x = L_1 + L_2$).}
\end{figure}

    Typical time dependencies of non-steady-state
amplitudes of the $+1$ and 0th diffracted orders are
presented in Fig. 2(a)--(d) at the resonant phase shifts
$\phi = \phi_r$ for the two different total grating widths:
$L = L_1 + L_2 = 2L_1 = 20\mu$m
($\phi_r \approx 191.2^\circ$---Fig. 2(a), (c)
and $L = L_c \approx 30\mu$m
($\phi_r \approx 205.2^\circ$---Fig. 2(b), (d);
recall that $\Delta L = 0$ (joint gratings). The curves in
Fig. 2(a), (c) have been calculated under the assumption
that the incident wave is switched on simultaneously
in the whole grating at $t = 0$ (see also [9]), while
the curves in Fig. 2(b), (d) have been calculated for
the incident wave being switched on at $t = 0$ everywhere
at the front grating boundary, and then propa-
gating through the grating. Both these approximations
are equivalent for non-steady-state DEAS (and EAS)
at times that are larger than the time that takes for the
incident beam to cross the grating. It can be seen that
these approximations give different results only for the
incident wave amplitude (compare Fig. 2(c) and (d))
and only at small time intervals ($\le 4\times 10^{-13}$s). Fast
oscillations of the curves in Fig. 2(d) within these
time intervals are associated with the computational
errors related to poor convergence of the Fourier
integral near the front end of the rectangular incident
beam (as it propagates through the grating). The sharp
jumps of the curves in Fig. 2(d) indicate the front end
of the incident beam passing through the middle of the
grating (dashed curve in Fig. 2(d)) and rear boundary
(solid curve in Fig. 2(d)). Note again that the
mentioned errors do not affect the results at larger times
after the front end of the incident beam has passed
through the grating. The effect of propagation of the
front end of the incident beam through the grating
on the non-steady-state scattered wave amplitudes is
completely negligible, because the scattered wave
amplitudes within these time intervals are close to zero
(Fig. 2(b)).

   One of the main results that can be derived from
Fig. 2(a)--(d) is the relaxation times for the
steady-state regime of DEAS in the considered gratings. It
can be seen that for all the curves in Fig. 2(a), (c) the
relaxation time is about $\tau \approx 6\times 10^{-9}$s. Within this
time interval, the scattered wave will propagate along
the grating the distance $\tau c\epsilon^{-1/2} \approx 80$\,cm. Taking
into account the angle of incidence $\theta_0 = 45^\circ$, this
distance gives the critical aperture of the incident beam
$l_c \approx \tau c\epsilon^{-1/2} \cos(\theta_0) \approx 57$\,cm.
Only if the aperture of
the incident beam is larger that the critical aperture
$l_c$, can the steady-state regime of DEAS be achieved
in the structure. It is obvious that for the considered
structure, the steady-state regime of DEAS is difficult
to achieve, since this would require excessively wide
coherent incident beams and, which is even more
problematic, a very long (longer than 80\,cm) grating.

   Therefore, we have considered non-steady-state
DEAS in a wider grating of $L = L_c \approx 30\mu$m ($\Delta L = 0$
and $L_1 = L_2$ )---Fig. 2(b), (d). In this case, as demon-
strated by Fig. 2(b), the relaxation time for the front
grating boundary is $\tau \approx 5\times 10^{-11}$s, while in the
middle of the grating and at its rear boundary  
$\tau\approx 2\times 10^{-10}$s. These relaxation times give the critical
apertures of the incident beam of $\approx 0.5$ and $\approx 2$\,cm,
respectively. These apertures (and thus the lengths of
the grating) are quite reasonable and easy to achieve
in practice.

    Another interesting feature that can be seen from
Fig. 2(a), (b) is that the relaxation at the front grating
boundary (dotted curves) occurs smoothly without any
noticeable bumps or minimums, whereas in the middle
of the grating and at its rear boundary the time
dependencies of the scattered and incident wave
amplitudes are characterised by a non-monotonic behaviour
(Fig. 2(a)--(d)). This suggests that the non-steady-state
$x$-dependencies of the incident and scattered waves
inside the grating should be non-symmetrical with
respect to the middle of the grating. This is indeed
demonstrated by Fig. 3(a), (b).

\begin{figure}[!htpb]
\centerline{\includegraphics[width=0.6\columnwidth]{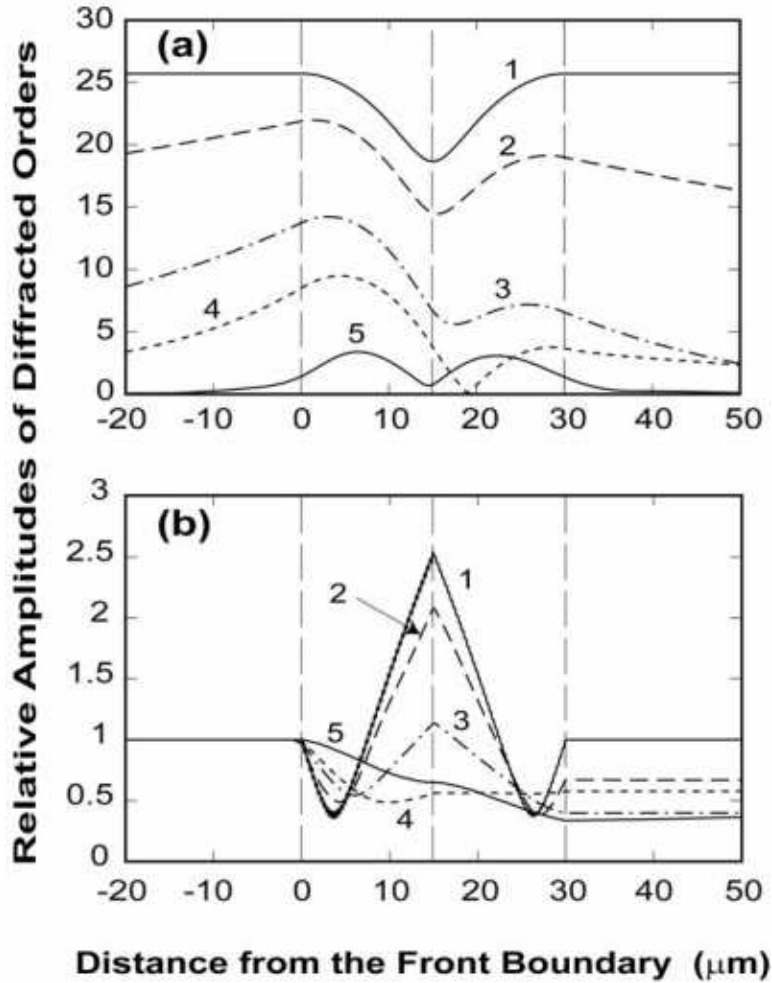}}
\caption{The dependencies of the relative scattered (a) and incident (b)
wave amplitudes inside and outside the two joint grating with DEAS
on distance from the front grating boundary at different moments of
time after switching the incident pulse on: (1) $t = \infty$ (steady-state
DEAS), (2) $t = 5\times 10^{-10}$s, (3) $t = 5 \times 10^{-11}$s,
(4) $t = 10^{-11}$s, (5) $t = 10^{-12}$s.
The structure is the same as for Fig. 2(b), (d):
$\epsilon = 5$, $\epsilon_{g1} = |\epsilon_{g2}| = 5\times 10^{-3}$,
$\theta_0 = 45^\circ$, $\lambda_0 = 1\mu$, $\Delta L = 0$,
$L_1 = L_2 = 15\mu$, $\phi = \phi_r \approx 205.2^\circ$.
The vertical dashed lines represent the grating boundaries.}
\end{figure}

    Curves 1 in Fig. 3(a), (b) represent the $x$-dependencies
of scattered (Fig. 3(a)) and incident (Fig. 3(b))
wave amplitudes for steady-state DEAS (see also
[1,2,4]) in the same structure as was used for Fig. 2(b), (d).
In particular, it can be seen that at small time intervals
(strongly non-steady-state DEAS) the amplitude of the
$x$-dependent scattered wave amplitude is characterised
by two distinct maximums in the middle of each of the
joint gratings (curve 5 in Fig. 3(a)). This is expected
because of the following reasons. Within small time
intervals (that are nevertheless larger than the time for
the incident pulse to cross the grating) the amplitude
of the incident wave is practically the same in both
the gratings (rescattering of a weak scattered wave
can hardly affect the amplitude of the incident wave
anywhere in the grating). Therefore, scattering of this
incident wave in both the joint gratings results in
the same amplitudes of the scattered waves. This is
the reason that there are two maxima on curve 5
in Fig. 3(a), approximately of the same height. At
the same time, due to the phase shift between the
joint gratings ($\phi = \phi_r \approx \pi$), the scattered waves
in these gratings will approximately be in antiphase
with each other. This is the reason that the scattered
wave amplitude is approximately zero at the boundary
between the gratings (curve 5 in Fig. 3(a)).

   With increase of time, the scattered waves start
diverging from one grating into another, and this leads
to the interaction between the joint gratings. As a
result, the minimum of the scattered wave amplitude
at the boundary between the gratings shifts to the right
(into the second grating), resulting in non-symmetric
dependencies (curves 2--4 in Fig. 3(a)). Due to this
process, both the amplitudes of the scattered and
incident waves in the grating increase, resulting in a
strong DEAS resonance (curves 1 in Fig. 3(a), (b)).
It is also interesting that once the maximum of
the incident wave amplitude in the middle of the
grating is established, it does not move anywhere from
the boundary between the gratings with increasing
time, but just increases in height---see curves 1--4 in
Fig. 3(b). For more detailed discussion of physical
reasons for DEAS in the considered structure see [1,2].

   It is also important to stress out again that for
the considered structures the approximate [1,2] and
rigorous [4] theories of steady-state DEAS give
approximately the same results (with the accuracy of
$\approx 7$\% for Fig. 2(a), (c) and $\approx 1$\% for Figs. 2(b), (d)
and 3(a), (b) [4]). Therefore, the presented dependencies
in Figs. 2 and 3 can simultaneously be regarded as
approximate and rigorous. The agreement between the
approximate and rigorous theories generally improves
with decreasing time intervals after switching the
incident wave on. This is because at smaller time
intervals the scattered wave amplitudes are smaller, which
makes it easier to satisfy the applicability conditions
for the approximate theory (see [10,14]).

   Previously [3], it has been demonstrated that strong
DEAS resonance can occur not only in two joint
gratings with a resonant phase shift, but also in two
uniform gratings separated by a gap of width $\Delta L \ne 0$
(Fig. 1). In this case, DEAS occurs due to diffractional
divergence of the scattered waves from one of the
gratings into another across the gap [3] (the mean
permittivity was regarded to be the same throughout
the structure).

\begin{figure}[htb]
\centerline{\includegraphics[width=0.6\columnwidth]{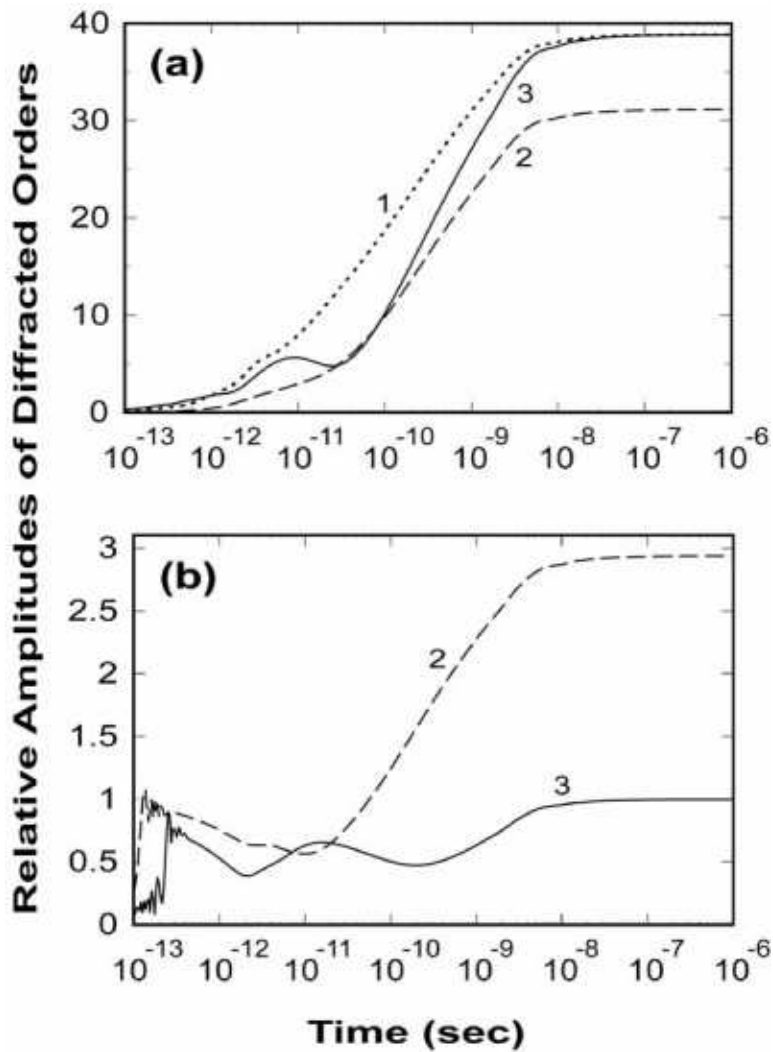}}
\caption{The typical time dependencies of the relative non-%
steady-state amplitudes of (a) the $+1$ diffracted order (scattered
wave) $|E_1/E_{00}|$, and (b) the 0th diffracted order (incident wave)
$|E_0/E_{00}|$ for DEAS in the two separated gratings with the
parameters:
$\epsilon = 5$, $\epsilon_{g1} = |\epsilon_{g2}| = 5\times 10^{-3}$,
$\theta_0 = 45^\circ$, $\lambda_0 = 1\mu$, $\Delta L = 10\mu$m
(separated gratings),
$L_1 = L_2 = 10\mu$, and $\phi = \phi_r \approx 213^\circ$.
Dotted curve---front boundary ($x = 0$),
dashed curves---middle of the gap
($x = L_1 + \Delta L/2 = 15\mu$m), solid
curves---rear boundary ($x = 2L_1 + \Delta L = 30\mu$m).}
\end{figure}

   The typical time dependencies of non-steady-state
amplitudes of the scattered and incident waves in the
middle of the gap (i.e., at $x = L_1 + \Delta L/2$) and at
the front and rear boundaries of the structure (i.e., at
$x = 0$ and $x = 2L_1 + \Delta L$---Fig. 1) are presented in
Fig. 4(a), (b). Here, we again assume equal widths
of the interacting gratings ($L_1 = L_2 = 10\mu$m). It can
be seen that there no significant differences between
Figs. 4(a), (b) and 2(a)--(d) (except for the lower
resonance in Fig. 4(a), (b) compared to that in the
same structure but without the gap--Fig. 2(a), (c)).
This clearly confirms that the physical reasons behind
DEAS in joint and separated gratings are the same---%
interaction between the gratings by means of the
diffractional divergence of the scattered wave from
one of the gratings into another [1--3].

   The approximation with the incident pulse being
switched on everywhere at the front grating boundary
at $t = 0$ has been used for Fig. 4(a), (b). This is
the reason for seeing the effects of propagation of
the incident wave through the grating at small time
intervals in Fig. 4(b) (similar to those in Fig. 2(d)).

   The relaxation times for the front boundary ($x = 0$),
middle of the gap ($x = L_1 + \Delta L/2$), and rear
boundary ($x = 2L_1 + \Delta L$) are
$\tau\approx 3\times 10^{-10}$s, $\tau\approx 6\times 10^{-10}$s,
and $\tau\approx 7\times 10^{-10}$s, respectively. These times
correspond to the critical apertures of the incident
beam $\approx 3$, $\approx 6$, and $\approx 7$\,cm,
respectively. Increasing
gap width between the gratings results in reducing the
height and sharpness of the DEAS resonance. This is
because the interaction between the gratings by means
of diffractional divergence becomes less efficient with
increasing gap width, since the scattered waves must
be spread by diffractional divergence across a larger
distance (for more detailed discussion see [3]). As a
result, increasing gap width results in decreasing
relaxation times in the gratings and the gap. Conversely,
decreasing gap width results in a rapid increase of height
and sharpness of the DEAS resonance [1--3]. Thus the
relaxation times in this case quickly increase, making
such a resonance difficult to achieve in practice (see
the comments for Fig. 2(a), (c)).

\begin{figure}[!t]
\centerline{\includegraphics[width=0.6\columnwidth]{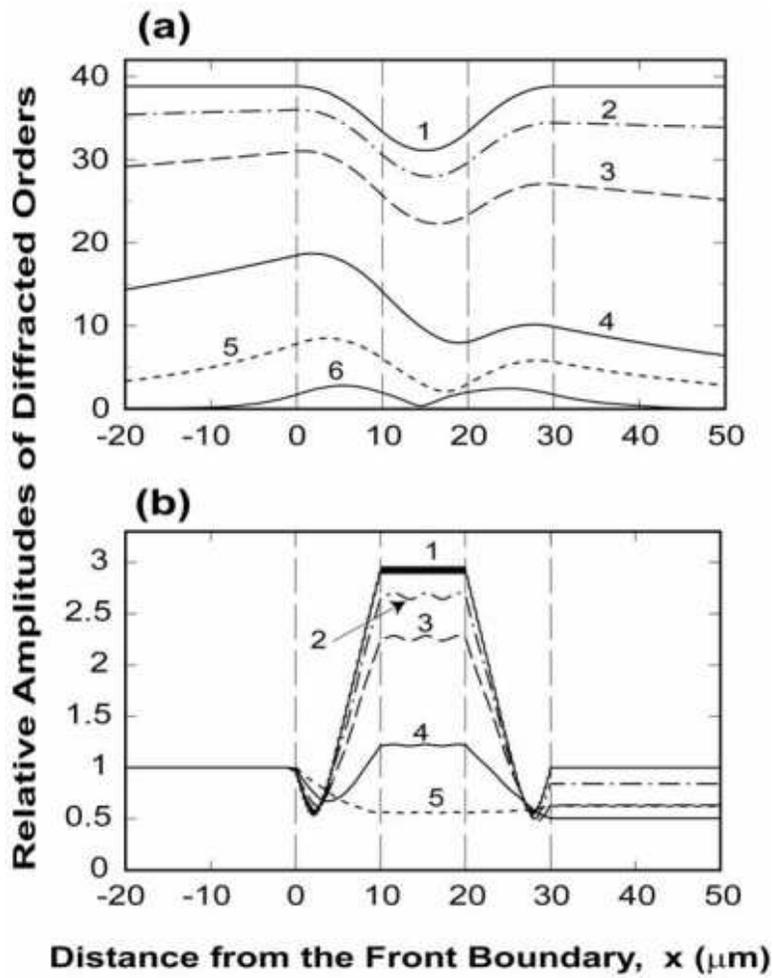}}
\caption{The $x$-dependencies of the relative scattered (a) and incident
(b) wave amplitudes inside and outside the two separated grating
with DEAS at different moments of time: (1) $t = \infty$ (steady-state
DEAS), (2) $t = 5\times 10^{-9}$s, (3) $t = 10^{-9}$s,
(4) $t = 10^{-10}$s, (5) $t = 10^{-11}$s, (6) $t = 10^{-12}$s.
The structure is the same as
for Fig. 4:
$\epsilon = 5$, $\epsilon_{g1} = |\epsilon_{g2}| = 5\times 10^{-3}$,
$\theta_0 = 45^\circ$, $\lambda_0 = 1\mu$, $\Delta L = 10\mu$m
(separated gratings),
$L_1 = L_2 = 10\mu$, and $\phi = \phi_r \approx 213^\circ$.
The vertical dashed lines represent the grating boundaries.}
\end{figure}

    The rigorous $x$-dependencies of non-steady-state
amplitudes of the scattered and incident waves in
the two gratings separated by a gap are presented in
Fig. 5(a), (b). In this case, at small time intervals, the
scattered wave amplitude is next to zero in the middle
of the gap (see curve 6 in Fig. 5(a)). This is again
expected, since the scattered waves with almost opposite
phases (due to $\phi = \phi_r \approx\pi$) diverge almost
symmetrically from the two gratings and practically
cancel each other in the middle of the gap. Further
increase of time results in the development of a more
complex pattern of interaction of the diverged waves
(that are not exactly in antiphase: $\phi_r\approx 213^\circ$),
eventually resulting in steady-state DEAS in the considered
structure (curves 1 in Fig. 5(a), (b)).

   The incident wave amplitude in the gap is constant
on average, but experiences small fast oscillations
(with the period of $\approx 0.3\mu$m---curve 1 in Fig. 5(b)).
As has been demonstrated in [10], these oscillations
are due to interference of the incident wave transmitted
through the first grating with waves resulting from
boundary scattering of the scattered wave at the
boundaries of the second grating. This effect is the
result of the rigorous theory of analysis of DEAS.
In the approximate theory, there is no boundary
scattering, and the amplitude of the incident wave in
the gap is constant [3]. Note that the discussed fast
oscillations can clearly be seen only on curve 1 for
steady-state DEAS. On the other hand, curves 2--5
display sinusoidal behaviour in the gap with much
larger spatial period (Fig. 5(b)). These oscillations in
fact must have the same period as those in curve 1;
the apparent longer period results from undersampling
and aliasing (i.e., insufficient number of points along
the $x$-axis). Note that the curves in Fig. 5(a) are
free from these computational errors, because they do
not contain fast oscillations that have to be resolved.
A similar (but less obvious) problem also occurred
in Fig. 3(b) (compare curve 1 with small and fast
oscillations and curves 2--5 where these oscillations
are not resolved--Fig. 3(b)).

   As has been mentioned above, if we disregard the
small and fast oscillations of the curves in Figs. 3(b)
and 5(b), the presented time and coordinate dependencies
in Figs. 2--5 can equally be regarded as rigorous
and approximate (this is the case for gratings with
sufficiently small amplitude [10,14]). On the other
hand, the approximate theory of steady-state DEAS
is immediately applicable for the analysis of scattering
of any types of waves (including bulk, guided and
surface optical and acoustic waves) in various kinds
of periodic gratings (e.g., periodic groove arrays for
guided and surface waves) [5,7,8,15]. Therefore, the
presented dependencies in Figs. 2--5 are also typical
for non-steady-state DEAS of any waves in gratings
with phase variations and sufficiently small amplitudes.

   The procedure of correlating the obtained dependencies
in Figs. 2--5 to the case of non-steady-state
DEAS of arbitrarily polarised guided slab modes in a
non-uniform periodic groove array is exactly the same
as that developed in [15] for steady-state EAS.
Obviously, if the grating amplitudes are large, then such
a simple correlation is no longer valid, and rigorous
theories for each type of waves and periodic gratings
should be developed.

\section{Conclusions}

   This Letter has carried out a detailed analysis of
non-steady-state DEAS in joint and separated gratings
by means of rigorous and approximate algorithms
based on the Fourier analysis of the incident pulse
and rigorous and approximate theories of steady-state
DEAS [1--4]. The main attention has been paid to the
investigation of non-steady-state DEAS of bulk TE
electromagnetic waves in holographic gratings with
step-like variations of phase. However, the obtained
results are directly applicable for all types of waves
in different kinds of periodic gratings with small
periodic modulation of structural parameters. This
statement is based on the universal applicability of the
approximate theory of steady-state DEAS in gratings
with sufficiently small amplitude [1--3].

    Typical relaxation times have been calculated for
several typical joint and separated gratings. The
corresponding critical apertures of the incident beam, that
are required for achieving steady-state DEAS, have
also been determined. In particular, it has been shown
that if the grating width is noticeably smaller than the
critical width [1,2], i.e., the DEAS resonance is
sufficiently strong, the relaxation time and the
corresponding critical aperture of the incident pulse can be too
large for this resonance to be achieved in practice. At
the same time, increasing grating width and/or introducing
a gap between the interacting gratings (separated gratings)
results in a quick reduction of height and
sharpness of the predicted resonance, and thus the
calculated relaxation times and critical apertures of the
incident beam. Typical structures that are reasonable
for practical observation of DEAS are thus identified.

\section*{References}

\begin{enumerate}
\item D.K. Gramotnev, D.F.P. Pile, Phys. Lett. A 253 (1999) 309.
\item D.K. Gramotnev, D.F.P. Pile, Opt. Quantum Electron. 32
    (2000) 1097.
\item D.K. Gramotnev, T.A. Nieminen, Opt. Quantum Electron. 33
    (2001) 1.
\item T.A. Nieminen, D.K. Gramotnev, unpublished.
\item M.P. Bakhturin, L.A. Chernozatonskii, D.K. Gramotnev, Appl.
     Opt. 34 (1995) 2692.
\item D.K. Gramotnev, J. Phys. D 30 (1997) 2056.
\item D.K. Gramotnev, Opt. Lett. 22 (1997) 1053.
\item D.K. Gramotnev, Phys. Lett. A 200 (1995) 184.
\item T.A. Nieminen, D.K. Gramotnev, Opt. Express 10 (2002) 268.
\item T.A. Nieminen, D.K. Gramotnev, Opt. Commun. 189 (2001)
     175.
\item M.G. Moharam, E.B. Grann, D.A. Pommet, T.K. Gaylord,
     J. Opt. Soc. Am. A12 (1995) 1068.
\item M.G. Moharam, D.A. Pommet, E.B. Grann, T.K. Gaylord,
     J. Opt. Soc. Am. A12 (1995) 1077.
\item D.K. Gramotnev, Frequency response of extremely asymmetrical
     scattering of electromagnetic waves in periodic gratings,
     in: 2000 Diffractive Optics and Micro-Optics (DOMO-2000),
     Quebec City, Canada, 2000, p. 165.
\item D.K. Gramotnev, Opt. Quantum Electron. 33 (2001) 253.
\item D.K. Gramotnev, T.A. Nieminen, T.A. Hopper, J. Mod. Opt. 49
     (2002) 1567
\end{enumerate}

\end{document}